\documentclass[twocolumn,fleqn,natbib,sort&compress,numbers]{svjour2}
\bibpunct{[}{]}{;}{n}{}{,} % to get "[numbered]" references from natbib
\smartqed  % flush right qed marks, e.g. at end of proof
\usepackage{graphicx}
\usepackage{amsmath}
\usepackage{epsfig}% Include figure files
\usepackage{dcolumn}% Align table columns on decimal point
\usepackage{bm}% bold math
\usepackage[latin1]{inputenc}
 
\journalname{Granular Matter}
\begin{document}

\title{Granular gases in mechanical engineering: on the 
origin of heterogeneous ultrasonic shot peening}

\subtitle{Granular gases in mechanical engineering}

%\titlerunning{Short form of title}        % if too long for running head
\author{M. Micoulaut$^1$, S. Mechkov$^1$, D. Retraint$^2$, P. Viot$^1$ and M. François$^2$}
\institute{$^1$Laboratoire de Physique Théorique de la Matière Condensée,
Universit{é} Pierre et Marie Curie, Boite 121,
4, Place Jussieu, 75252 Paris Cedex 05, France\\
$^2$ Laboratoire des Systèmes Mécaniques et d'Ingénierie Simultanée, 
Université de Technologie de Troyes, BP 2060 -10010 Troyes Cedex, France}

\date{Received: date }
% The correct date will be entered by the editor

\maketitle
\begin{abstract}
The behavior  of an ultrasonic  shot peening  process is  observed and
analyzed by using a model of inelastic hard spheres in a gravitational
field that are  fluidized by a vibrating  bottom wall (sonotrode) in a
cylindrical chamber.  A  marked heterogeneous  distribution of impacts
appears when the collision between the shot  and the side wall becomes
inelastic  with constant dissipation.    This effect  is one order  of
magnitude larger than  the simple heterogeneity arising  from boundary
collision on the cylinder. Variable restitution coefficients bring the
simulation closer   to    the  general observation  and    allows  the
investigation of  peening regimes with  changing shot density. We
compute  within   this   model  other  physical   quantities   (impact
velocities, impact angle, temperature  and  density profile) that  are
influenced by the number $N$ of spheres.
\end{abstract}
\maketitle
\section{Introduction}
It  is    well known that   the introduction   of residual compressive
stresses  in metallic components   leads  to reduce fatigue   strength
\cite{FS80}.  Therefore, many  engineering techniques involve surface
treatment to allow  either  surface hardening  (by e.g.   nitriding or
vapor deposition) or   fatigue life improvement \cite{CRBE99}  through
laser shock peening  or shot peening. For the  latter, a high velocity
stream of steel particles is projected at a material surface producing
at and below it compressive residual  stresses with a peak value being
reached  at some depth  below the  surface \cite{FBC96}.  A particular
mechanical treatment derived from  conventional shot peening is called
{\em  ultrasonic shot  peening},   ultrasonic is a  reference  to  the
frequency  of vibration of the  sonotrode (see below). It has received
attention in the  recent  years  \cite{LPNBF94} since  it could  be  a
promising  technique  for  obtaining  surface  treatment of   metallic
surfaces.  

Here, a piezoelectric generator produces the  vibration of a sonotrode
that projects upon contact  steel shot in a  chamber closed by a cover
which is the sample to be  peened.  The shot  is usually made of small
steel particles whose  diameter is  between   $1$ and  $3~mm$ and  the
frequency is  about $20kHz$.  Several  parameters can also be changed,
allowing one to control of the overall  shot velocity and thus the shot
peening intensity. Possible tuning  parameters for optimizing
the peening process include the shot  diameter, the height of
the  chamber, the  amplitude  and/or the frequency  of the  sonotrode.
Basic applications  of  this technology  are   found in  automotive or
aerospatial industry.

If  the performance  of  this   process is   closely related  to  the
appropriate choice of  parameters, it becomes necessary  to understand
how   the peening intensity  or  the peening   distribution on a given
sample  is affected    by    changes  of mechanical or      electrical
characteristics of  the system.  Furthermore,  as some of the physical
quantities  involved in the peening  process  are hard \footnote{ Here, it is a real-time experiment.} to
measure  in a real-time  experiment (velocity, acceleration,\ldots), mainly
for safety  reasons (the impact of  steel beads  are so strong
that the  chamber is  a  closed box),  numerical   simulation can  be a
powerful  tool for investigating   the  influence of these  quantities
under various situations.  

In parallel, there has been great activity during 
the last ten years in the study
of granular gases \cite{GG01,PB03,G05}.   In particular, systems of
vibro-fluidized glass beads   in an cylinder has some  similarities
with the device used for the shot peening \cite{WHP01,WHP01a}. The main
difference  with  the experimental setup  used in ultrasonic 
shot peening is the
absence of the cover and a lower frequency of vibration. In the latter
series  of experiments, it was shown  that  the inelastic sphere model
provides    an    accurate        description  of          microscopic
quantities \cite{TV02,TV02a} (local  granular temperature,  local  mean
velocity, local density, ...)  which  encourages us  to perform a
molecular dynamics of inelastic hard spheres.

The  system is represented  by a collection  of inelastic hard spheres
colliding with each other and with the  boundaries (chamber, the shot,
and    sonotrode).  For the sake   of    simplicity, collisions are  first
characterized  by a constant  normal  coefficient of restitution.   We
perform an event-driven Molecular Dynamics that is as close as possible to
the experimental  setup  by using   the geometrical   features of  the
chamber,  sonotrode  and the  cover.  In  order  to  obtain an improved
description  of  the  model, we also
consider  a model of inelastic hard  spheres where the restitution
coefficient depends on the relative velocity of the impact.

Our  results both theoretical and  experimental show  that the peening
distribution  on the sample is not  homogeneous.  The heterogeneity of
the peening  distribution is strongly influenced  by  the value of the
particle-side wall coefficient of restitution $c_w$.  This result goes
far beyond the intuitive view  that heterogeneity should simply result
from the boundary collisions  on the side  walls. The increased energy
dissipation along the   side walls favors  particle  accumulation thus
increasing the  gas (shot) density on the  border of the chamber. This
leads  to an increase  of the  impact  frequency on  the border of the
sample.  Within the model, we compute impact velocity and impact angle
of the shot and show  also a changing behavior  with the shot density,
ranging from the dilute Knudsen limit to a more dense situation where
inter-particle collisions dominate.   Both  quantities  display marked
differences  between  the  border  and the    center of  the top  wall
(sample).
\section{Inelastic hard sphere model}

\subsection{Simulation details}

We first  consider  the model close   to  the experimental  setup (see
below).  The cylinder has a radius of $R=35~mm$ which contains $N=200$
hard  spheres representing the shot of  diameter $3mm$. The latter are
subject to a constant gravitational force.  The  energy is supplied by
vibrating the bottom wall following a symmetric saw-tooth profile with
amplitude $A$   and  period $T$   which mimics in   the simulation the
sinusoidal profile  of the sonotrode \footnote{  By  using a saw-tooth
profile the  time of a collision  between  a particle and  the base is
obtained analytically ,   whereas   with a sinusoidal  profile,    the
collision time is given by an implicit  equation which requires a more
expensive numerical computation.}.  One should note that the choice of
this profile has no major impact on  the results\cite{MB97}, since the
amplitude of the harmonic  $n$ of the saw-tooth  profile falls as  odd
$n^{-2}$.  We mention also that even  though the electrical excitation
of the sonotrode is sinusoidal,  because of the elastic deformation of
the  sonotrode, the  velocity  applied to  the  shot is  certainly not
purely sinusoidal \footnote{There can be a  difference in amplitude of
the sonotrode on the border and on the center.}.

The spheres collide inelastically and instantaneously with each other,
with the cylindrical side   walls, with the  top   wall and with   the
sonotrode. The  corresponding constant coefficients of restitution are
denoted  $c$, $c_w$, $c_b$ and $c_t$.   The different collision rules   are
given by the following expressions:
\begin{eqnarray}
\label{cw}
{\bf v}'_{i,r}={\bf v}_{i,r}-(1+c_w)({\bf v}_{i,r}.\hat {\bf
r}_{i,r})\hat {\bf r}_{i,r}
\end{eqnarray}
\begin{eqnarray}
\label{cb}
v'_{i,z}=v_{i,z}-(1+c_b)(v_{i,z}-v_S)
\end{eqnarray}
\begin{eqnarray}
\label{ct}
v'_{i,z}=-(1+c_t)v_{i,z}
\end{eqnarray}
\begin{eqnarray}
\label{c}
{\bf v}'_{i,j}={\bf v}_{i,j}±{\frac {1+c}{2}}[({\bf v}_j-{\bf
v}_i). \hat {\bf n}]\hat {\bf n}
\end{eqnarray}
where the  prime  quantities denote  the  post-collisional quantities;
${\bf v}_{i,r}$ and  $\hat {\bf r}_i$  is the unit  position vector of
particle $i$ are the velocity and the  position of the particle $i$ in
the  horizontal plane  respectively and   $c_w$  the normal coefficient  of
restitution for a collision between a sphere and the chamber. 
The   particle-bottom (sonotrode) wall
restitution coefficient $c_b$ is first taken as  
unity which amounts to rescaling
the amplitude of the  vibration. $v_S$ is
the vertical velocity  of the sonotrode   ; $v_{i,z}$ is the  vertical
component  of the   velocity of  particle $i$  and  $c_t$  the  normal
coefficient of  restitution for a collision between   a sphere and the
cover; finally, ${\bf v}'_{i,j}$  denote the velocities  of $i$ or $j$
particle,  $\hat {\bf n}$ is  the unit center-to-center vector between
the colliding  pair $i$ and  $j$ and $c$ is  the normal coefficient of
restitution for a sphere-sphere collision.
One should note that there are two rules along the $z$ axis (equations
(\ref{cb}) and (\ref{ct}))  depending on    which wall  the    spheres
collide: bottom (sonotrode) or top  (sample).  Between collisions, the
spheres    follow   parabolic  trajectories   due   to   the  constant
gravitational  field (viscous damping  with  the air  contained in the
chamber is neglected).
\par
\begin{figure}
\begin{center}

\resizebox{7cm}{!}{\includegraphics{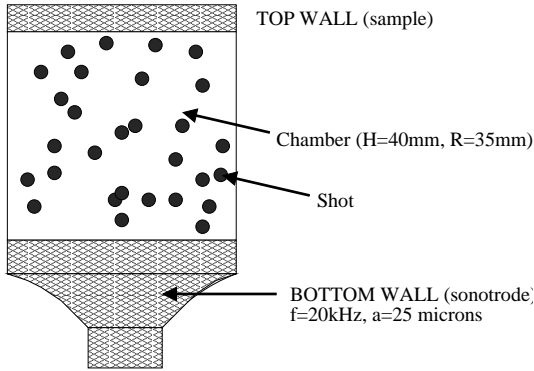}}
\end{center}
\caption{\label{chambre} Sketch of the  experimental setup that is used as simulation box
in the inelastic hard sphere model.}
\end{figure}
The parameters of the  model used in  the simulation are obtained from
experiment (Figure \ref{chambre}). The vibration  frequency   is $20kHz$ and the
amplitude  of the   sonotrode  is  $25\mu  m$.   The  chamber height  is
$40~mm$. In the first part, we carry out  most of the simulations with
the restitution  coefficients  $c=c_t=0.91$ corresponding  to the usual
experimental values for steel shot in the velocity range of interest
\cite{Ecolemines}  and leaving as an adjustable  parameter  only the side
wall restitution  coefficient $c_w$.  This is in order to  highlight the
strong  influence   of   the  side-wall   collisions on   the   impact
heterogeneity.   We stress however  that the overall observed behavior
of the model does not depend on the precise values given for $c_t$ and
$c$ and several additional runs with different $c_t$ and $c$ have been
performed to check the robustness  of our conclusions.  The stationary
non-equilibrium  state  is  achieved by   a  preliminary simulation of
typically $5000$ collisions per particle which  corresponds to a peening
time of
about $1ms$.  The collision time estimated as $1\mu s$ from Hertz theory
\cite{G90} is much lower.  The  statistical analysis of the quantities
of interest  has been accomplished  for a  total simulation  time of
$26~s$ corresponding to $5×10^6$ collisions.
\subsection{Velocity dependent restitution coefficient}
In a  second  part  (section~\ref{sec:comp-with-observ}),  the present
model is made more realistic by taking into account velocity dependent
normal  restitution coefficients  that  depend  on the normal   impact
velocity.       This       dependence     is        rather        well
known \cite{R18,T48,O22,A30,LRD97}  and was  first reported in the
1920's.  Plastification  under  high velocities  (typically  when  $v\geq
5~m.s^{-1}$) has also been   reported.  In the high  velocity   limit,
experimental measurements suggest \cite{G90} a  power-law  behavior of the  form:
$c_N\propto v^{-1/4}$ whereas in  the  low velocity range, the  deformations
are   supposed  to   be   elastic    and  dissipation    described  by
visco-elasticity \cite{HSB95,BSHP96}.  For   the  latter,   it has been
obtained \cite{HSB95} a slightly different power-law  which is like    
$(1-c_N)\propto v^{-1/5}$.

The  behavior of restitution coefficients  with respect to some easily
measurable parameter is, however, a  much deeper problem that can
certainly not be   encoded in  the simple aforementioned   power-laws.
Experiments have indeed shown that  $c_N$  could depend on the  sphere
density \cite{J85},  the  sphere diameter   or   the thickness of   the
impacted       surface \cite{Z41},     or     even      the      impact
angle \cite{RCB90,HGS99}.   Beyond  the details  of a  given  material,
studies on the restitution coefficient all suggest a generic threshold
between  a regime at low   velocity  and low  dissipation where  $c_N$
depends  weakly \cite{BP03,SP98}  on the impact  velocity,   and a more
dissipative     regime    induced     by     plasticity    (or   even
fracturing) \cite{KG00,T97}.

Recently,  simulations of  inelastic hard spheres  have been performed
using  variable   restitution   coefficients \cite{MF05}   and    have
shown the necessity of the latter to accurately describe
experiments.    Specifically, pressure  effects as a   function of the
density  of spheres could be recovered  by simulation for a dilute and
dense  vibrated  granular medium.  It   has been shown also that  the
unphysical clustering tendency  was  reduced with the use  of velocity
dependent restitution coefficients.

In  section IIIB,  we will use    a threshold model   for the normal
restitution coefficient defined by:
\begin{eqnarray} 
c_N^i(v) = \left\{ \begin{matrix} 
c_0^i, &  v \leq v_0^i, \cr 
c_0^i \left( \frac{v}{v_0^i} \right)^{-1/4} & v \geq v_0^i \end{matrix}\right. 
\label{cvar} 
\end{eqnarray} 
where $v_0^i$ is a threshold velocity, $c_0^i$  is the constant normal
restitution coefficient at  low velocity and $i=b,t,w,s$ following the
nature of    the   impacted surface   (bottom,   top,  wall, spheres).
Parameters are  given in Table I. One  expects indeed that softer
materials such as the aluminum side walls or the sample will have a lower
threshold  velocity and  a lower  $c_0$   as compared to the  titanium
sonotrode or the steel spheres.
\begin{table}
\begin{centering}
\begin{tabular}{lcccll}
\hline\hline
Impacted material& &i& &$c_0^i$&$v_0^i$ [cm/s] \\
\hline
Sonotrode (titanium)& &b& &0.91&1.2 \\
Spheres (steel)& &s& &0.91&1.2 \\
Sample (aluminum)& &t& &0.6&0.12 \\
Side walls (aluminum)& &w& &0.6&0.12 \\
\hline\hline
\end{tabular}
\end{centering}
\caption{Parameters for the inelastic hard sphere model with variable 
restitution coefficient.}
\end{table}
\par
Finally, we also use the simplest possible  model to take into account
the  transverse dissipation that  leads   to a tangential  restitution
coefficient $c_T$. To our knowledge, only a very few studies have been
reported on the subject (see however \cite{B88,GNB05}).  The total loss of
translational kinetic energy   can usually be  described  by the total
restitution   coefficient $c=[c_N^2\cos^2\theta+c_T^2\sin^2\theta]^{1/2}$  where
$\theta$ is the impact  angle. Conservation of impulse  and momentum and an
additional  condition of rolling prior to  departure from the impacted
surface leads to a value of $c_T=5/7$ (the factor $5/7$ comes from the
momentum of intertia). Here, it is assumed that the loss in kinetic energy
mostly arises from the sphere rotation during the impact. One may also
assume that slip continues throughout  contact which will in this case
lead to:  $c_T=1-\mu(1+c_N)\cot\theta$. But this   would define an additional
parameter $\mu$ corresponding  to the Coulombic friction coefficient  of
the impacted surface.   Experimental  measurements  of steel   spheres
bouncing on   flat aluminum plates   show \cite{RCB90} that the constant  value of
$c_T=5/7$  is mostly valid at  small  impact angles  and under certain
conditions up to  $\theta \simeq 55^o$. However, a  more  detailed analysis that
should include the  deformations  and velocities associated  with  the
elastic deformations  of the surfaces is  clearly beyond the scope and
objectives of this paper. Since we are handling instantaneous impacts,
effects of friction or peculiar material  properties can only be taken
into account   via an  {\em  effective}  velocity dependence  of  the
restitution coefficients.  Furthermore, we  stress that this will  not
affect the general observed behavior with shot density.

\section{Results}
\subsection{Origin of the impact heterogeneity}
\begin{figure}
\begin{center}
\resizebox{7cm}{!}{\includegraphics{compar.eps}}\\
\resizebox{7cm}{!}{\includegraphics{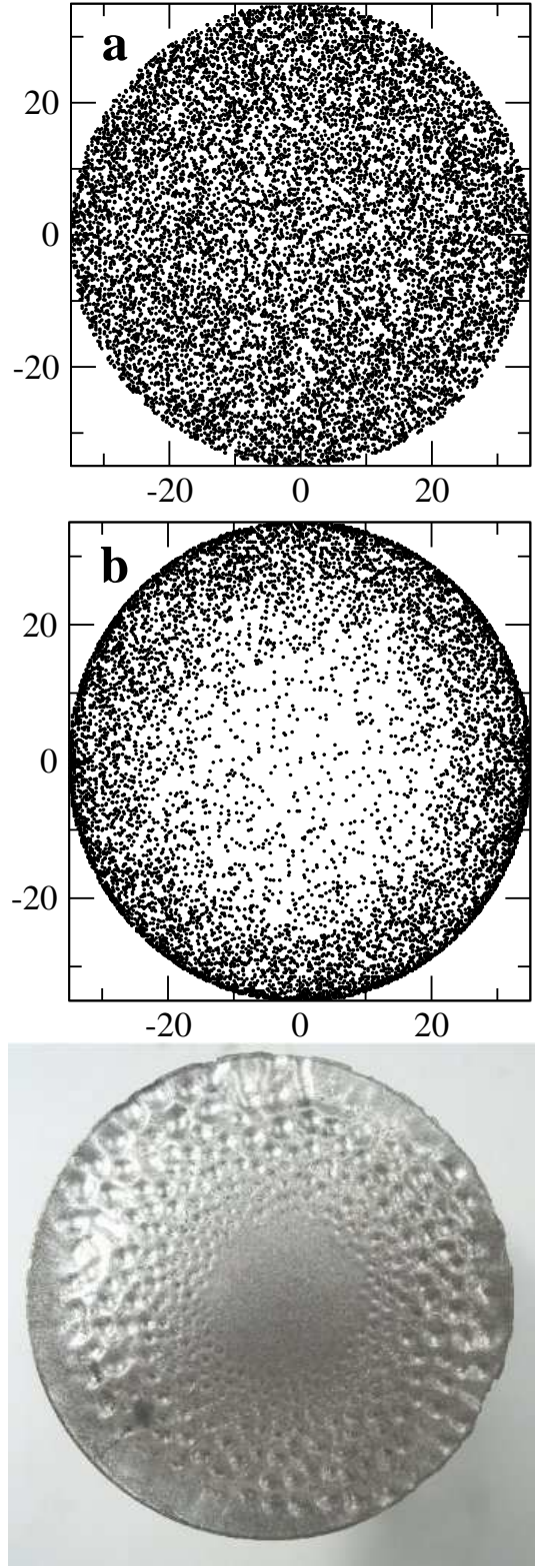}}

\caption{\label{compar} Distribution of impacts on the top wall (sample).
a) $c_w=0.91$. b) $c_w=0.20$. c) Sonotrode after several hours of use. See text 
for details.}
\end{center}
\end{figure}

Figure \ref{compar} shows the different impact profiles that appear on
the  top wall of  the  chamber  after  $1~s$  simulation time  for two
different values of the side wall  restitution coefficient $c_w$.  For
$c_w=c=c_t=0.91$,   one  has  an  almost  homogeneous distribution  of
impacts  (Fig \ref{compar}a) whereas  heterogeneity sets in when $c_w$
is lowered to $0.20$ (Figure  \ref{compar}b). The present results have
to  be contrasted  with the observed   profiles obtained on the impacted
aluminium sample after $1~ s$ or on the sonotrode.  It clearly shows
that the inelastic sphere collision with the side wall are relevant
for understanding the heterogeneous shot peening which is 
manifested by an increased impact
number on the border of the sample.  Nonelastic collisions on the side
wall originate the impact profile that  is experimentally observed (e.g.
on the sonotrode  (bottom wall), Fig \ref{compar}c) and recovered from
the simulation.  The same profiles are  obtained for the top wall with
a similar impact frequency (see below).
\par
\begin{figure}
\begin{center}
\resizebox{7cm}{!}{\includegraphics{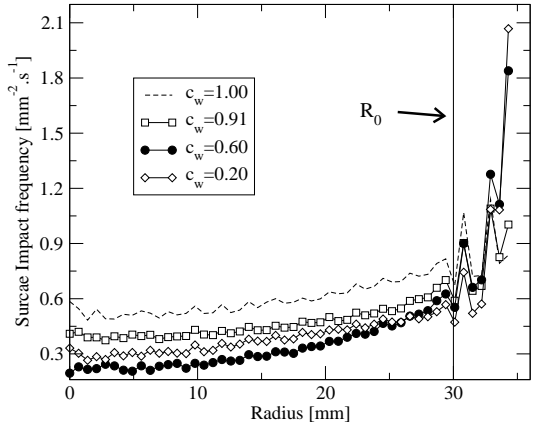}}

\end{center}
\caption{\label{nsurf}Surface impact frequency with respect to the radial 
distance $R$ for different side wall restitution coefficients $c_w$ (symbols). 
The broken line corresponds to the pure elastic side wall collision 
($c_w=1$). The 
distance $R_0$ is used below (see text).}
\end{figure}
The  influence of   the   sphere-wall coefficient  of  restitution  is
quantitatively  observed by monitoring  the  impact frequency   per
 surface unit ${\cal N}$ with respect to the  radius $R$ of the chamber
(Fig.~\ref{nsurf}).  It results that when  the shot has a pure elastic
collision on  the side walls  ($c_w=1.0$), then the  number of surface
impacts slightly   increases from the   center to  the radius  of  the
chamber,    whose values are    in     the range ${\cal    N}=0.6-0.8$
impact$.mm^{-2}.s^{-1}$.  When $c_w$ decreases, a similar behavior is observed
when the radial  distance is lower  than a typical  value $R_0\simeq30~mm$.
For $R>R_0$ significant differences  appear, e.g., for $c_w=0.60$, the
number  of surface impacts   is  multiplied by  almost  a factor  $10$
between the center and the  border of the  cover and this increase  is
even more dramatic for the ultimate value of $c_w=0.20$.

Intuitively, one  expects that the     impact distribution should   be
heterogeneous even in the case of elastic side wall collisions because
of the lateral bounce on the cylinder wall.  The simulation shows that
this effect clearly seen from Fig.  \ref{nsurf} when $c_w=1.0$ is weak
as compared to the effect arising from the decrease of $c_w$. Boundary
collisions  produce  an  increase    of  the  surface impacts     from
$\simeq0.5~mm^{-2}.s^{-1}$ in the  center to  about $0.8~mm^{-2}.s^{-1}$ on
the border  for  elastic side wall collisions  whereas   the effect of
lower  side wall restitution   coefficients (e.g.  $c_w=0.6$) leads to
${\cal N}=1.8~mm^{-2}.s^{-1}$ on  the  border  of the top  wall, 
compared to ${\cal N}=0.25~mm^{-2}.s^{-1}$ on the center.   The
present  results do not depend crucially  on  the values taken for the
other restitution  coefficients  as  similar trends for    the surface
impact frequency are obtained for lowered $c$ and $c_t$. For instance,
when $c=c_t=c_w=0.6$, the  trend with $R$  observed is rather close to
the one  displayed  in Fig.   \ref{nsurf}  with $c_w=0.6$, except that
${\cal N}$ ranges  now from $0.1$  to $1.7~mm^{-2}.s^{-1}$ between the
center and the border of the sample.

Beyond the numerical details,  the origin of the heterogeneity becomes
clear. With increased dissipation on the side  walls, the spheres have
a reduced velocity and are ``adsorbed'' on the  side walls with an upward
helicoidal trajectory arising from the impulse of  the sonotrode. As a
result, the density  and the granular  temperature (kinetic energy per
particle) of the granular gas appear to be also strongly influenced by
the dissipation.

\subsection{Comparison with observation}\label{sec:comp-with-observ}

Aluminum   samples have  been  peened  during  1  sec   and the  impact
distribution  displays a   weak  heterogeneity. The average  roughness
between the  border and the center  of the sample is respectively $4~\mu
m$  and $7.4~\mu m$. In general, the obtained roughness reveals the degree
of impact on a sample. It provides therefore an indirect evidence that 
more shot  has been impacting the border. The 
heterogeneity can be quantified  in a  manner similar to
Fig.\ref{nsurf} by sampling circularly the frequency of impacts  $\cal
N$ per unit surface with respect to the radius of the sample. $\cal N$
varies from $0.65$ at the center of  the sample to
$0.95$  on the border.  The same  tendency is  obtained with rectangular
sampling, i.e. one obtains ${\cal N}=1.25 ± 0.12$ at the center of the
sample,  and ${\cal N}=  1.66 ± 0.39$  at the border.  The presence of
polymer adhesive stripes  on the side  walls that induce an  increased
dissipation, leads to a lowering of the respective impact frequencies,
in harmony  with the observed   trends displayed in Fig.  \ref{nsurf}.
Refinement of the measurements is currently under consideration.

\subsection{Selective shot peening}
Once the origin of the  heterogeneity is identified, we investigate the
effect of the  sphere density (or sphere number  $N$) in  the chamber on
the  peening statistics. Within  the inelastic hard sphere model using
now the "{\em realistic}"  restitution coefficients  of Eq. (\ref{cvar}),
one has indeed  the luxury to investigate  features appearing with the
progressive   jamming  of the system,  starting   from the very dilute
limit, and characterize the deeper origin of the obtained profiles.

\begin{figure}
\begin{center}
\resizebox{7cm}{!}{\includegraphics{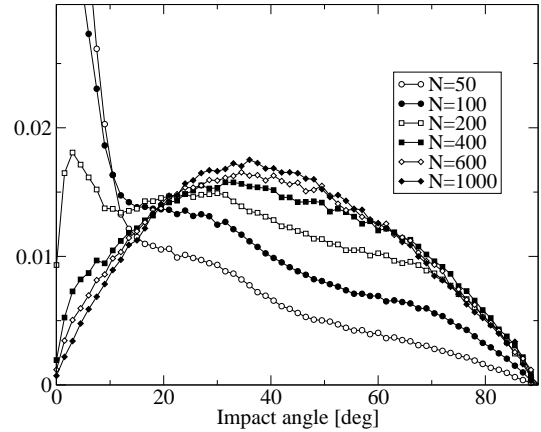}}
\end{center}
\caption{\label{angletop} Impact angle distribution on the top wall for
 different number of spheres $N$. All distributions are normalized to one.}
\end{figure}

\begin{figure}
\begin{center}
\resizebox{7cm}{!}{\includegraphics{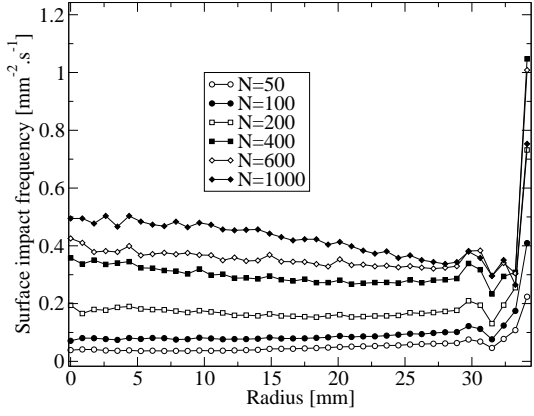}}
\end{center}
\caption{\label{nsurf1} Surface impact frequency $\cal N$ on the top wall as a 
function of the radius of the chamber for different number of spheres.}
\end{figure}

Figure \ref{angletop} shows the   impact angle distribution on  the top
wall (the sample) for various number of  spheres. One can observe that
the nature of the  impacts can be  very  different following  the very
dilute  ($N=50$,  density $\eta_0=3.25×10^{-4}~mm^{-3}$)   or  more dense
($N=1000$, $\eta_0= 65.0×10^{-4}~mm^{-3}$) situation. At low density, the
impact is almost normal with a very sharp distribution centered around
the impact angle  $\theta=0^o$. Here one sees  that the spheres will mostly
bounce back  and  forth between the  top  and the  bottom walls with a
rather  small number of inter-particle  collisions.   As a result, the
surface impact frequency  $\cal N$ with respect   to the radius  (Fig.
\ref{nsurf1}) is rather flat and starts only to grow close to the side
walls.  With increasing $N$, these simple (mostly linear) trajectories
tend  to disappear as more and   more sphere-sphere collisions are now
involved. Finally, the distribution becomes very broad at high $N$ and
centered around $\theta=35^o$ and there  is not much difference between the
system with  $N=600$ and $N=1000$ spheres.   Additional spheres do not
change  the obtained distribution.  This shows  also  that in the more
dense  situation,    normal  impacts    are very  rare   on  the   top
wall, i.e. the probability of having $\theta=0$ is almost zero. 
Noteworthy is the system with $N=200$ spheres which displays
an angular distribution that contains a reminiscent signature 
of the very dilute limit,
i.e.  showing a shallow peak around $\theta=5^o$  which disappears when $N$
is   increased from $200$   to $400$. It suggests  that in this intermediate 
situation, some spheres
succeed  in moving upwards   through  the  granular  gas without   any
sphere-sphere collision.

Figure  \ref{nsurf1} shows an  additional interesting feature which is
the occurrence of  a  local order  produced   by the accumulation of the
spheres on the side   walls.  With increasing  $N$, more and  more spheres are 
trapped on the side walls, similarly to the result of the constant restitution
coefficient model (Figure  \ref{nsurf}).  This produces an  increased
jamming in the  vicinity of the side wall  which does not allow  other
arriving spheres  to reach it.   As a result,  these spheres will stay at a
distance of the order  of the diameter   $\sigma$ of the spheres.   This is
reflected in the quantity $\cal N$  by an impact  frequency peak at  about $R=30~mm$
and even a secondary peak   for high densities ($N=600$ and  $N=1000$)
between the latter value and the border of the sample.

Finally, we note that for a large number of spheres ($N=1000$),
the  surface impact frequency decreases from  the center of the sample
to the first impact frequency peak at $R=30~mm$.  
\begin{figure}
\begin{center}
\resizebox{7cm}{!}{\includegraphics{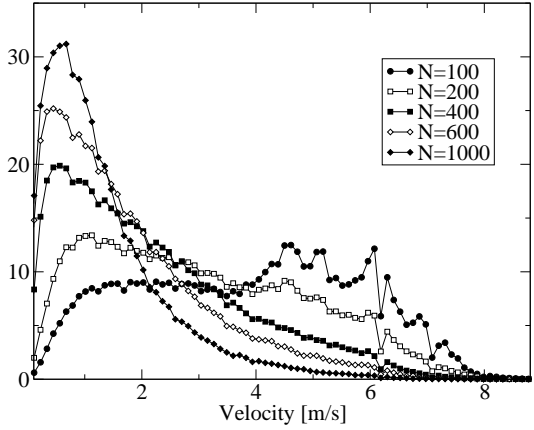}}
\end{center}
\caption{\label{topvit} Vertical impact velocity distribution on the top wall 
for different number of spheres N. All distributions are normalized to one.}
\end{figure} 
With the increased number of spheres, the impact velocity distribution
(Fig. \ref{topvit}) refines and  converges to a Maxwell-Boltzmann-like
distribution  that  can be  fitted   by $v_z\exp(-v_z^2)$ with  a mean
velocity that is of about $0.8~m/s$ and which is close to the measured
velocities for  this  kind of system \cite{Ecolemines}.  
Note  that for  a small
number of  spheres, the  impact   velocity distribution is   broad and
ranges from $1~m/s$ up to large velocities  of about $6~m/s$ corresponding
to  spheres that have  been optimally accelerated  by  the sonotrode.  
Once  the
system densifies, the dissipation due to sphere-sphere   collisions
lowers the overall impact velocities.

\begin{figure}
\begin{center}
\resizebox{7cm}{!}{\includegraphics{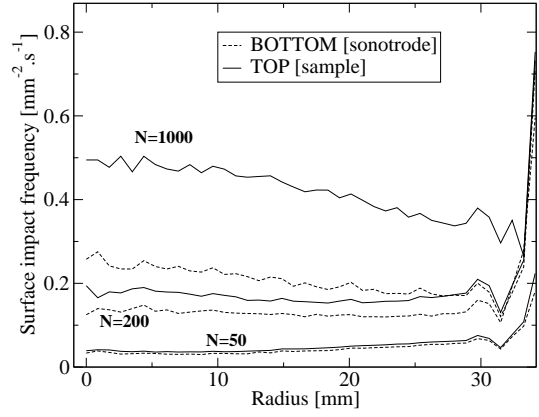}}

\end{center}
\caption{\label{nsurfbotop} Surface impact frequency $\cal N$ on the top 
(solid lines, same as Fig. \ref{nsurf}) 
and the bottom (broken lines) walls as a function of the radius of the chamber.}
\end{figure}

\begin{figure}
\begin{center}
\resizebox{7cm}{!}{\includegraphics{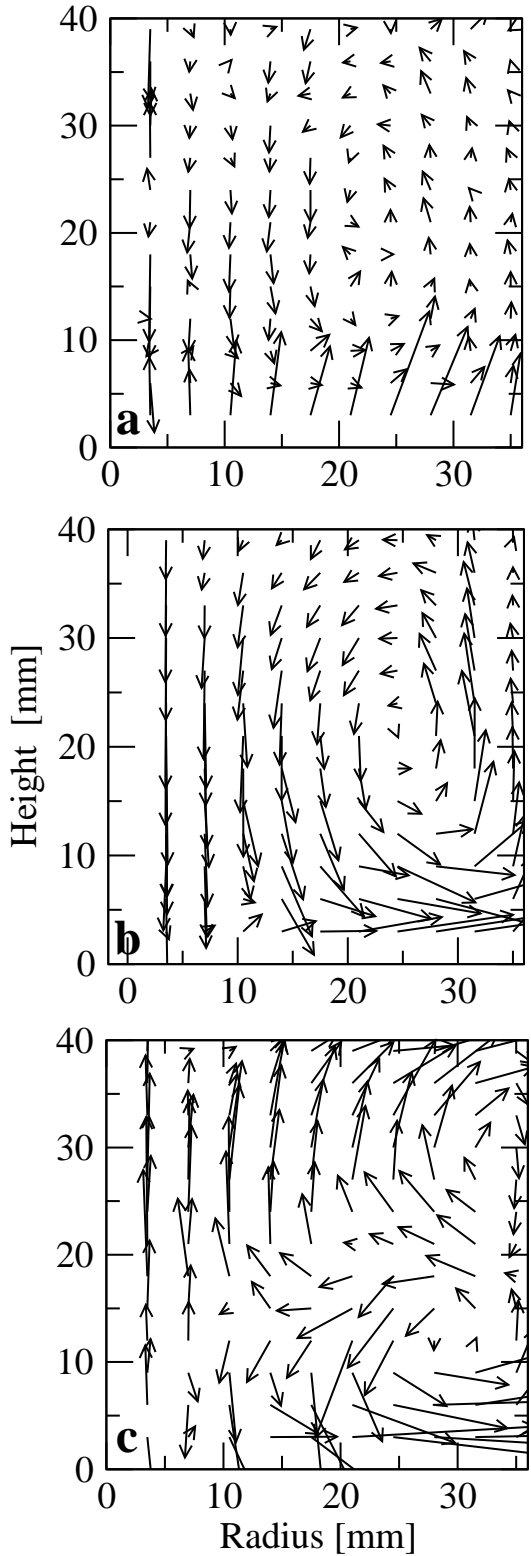}}

\end{center}
\caption{\label{chamvit} Mean velocity field of the inelastic hard spheres 
in the (r,z) plane for three different numbers of spheres. a) N=50, b) N=200, c) N=1000.}
\end{figure}

A comparison between the  surface impact frequency of  the top and the
bottom  walls  (Figure  \ref{nsurfbotop})  shows that the
impact regime can  be rather  different  but still consistent  with previous
findings. For a very small number  of  spheres ($N=50$),  the result on  the
surface impact frequency  is  consistent with our  previous  findings,
i.e.  $\cal   N$  is identical  between the   top and the   bottom, in
agreement with the conclusion drawn from the impact angle distribution
(Fig.
\ref{angletop}), i.e. quasi-normal trajectories for the spheres which 
collide almost only between bottom and top. For an increased number of
spheres,   differences  emerge which   can  become very  pronounced as
suggested by the value of the impact frequency on the center of sample
for   $N=1000$.     For   the  top   wall,    it     is found   ${\cal
N}=0.50~mm^2.s^{-1}$   whereas   ${\cal   N}=0.26~mm^2.s^{-1}$ for the
bottom wall, i.e.  there  is almost a  factor  of two between the  two
colliding walls.
In order  to infer the  origin of  the peening  difference  of the two
surfaces, we have computed  the mean velocity  field for the system of
inelastic spheres in the ($R,z$) plane. The field is averaged over the
azimuthal  angle  and  $5×  10^6$ collisions.  Figure \ref{chamvit}
shows the velocity field for the corresponding  number of spheres used
in Fig. \ref{nsurfbotop}.
For $N=200$,  a toroidal convection roll is  clearly present in which
the particles  flow, on average,   up from the    border and down  the
center.  This kind  of    convection roll has    been  found  both  in
simulations and experiment for open vibrated granular media
\cite{WHP01,TV02} at vibration frequencies of $50~Hz$. The convection
roll is  maintained  (but weaker by  about an  order of   magnitude in
intensity) for  a lower number ($N=50$) of  spheres. However, one sees
that  for  $N=1000$,  the convection  roll   breaks up  and
 spheres flow from the side wall either  upwards to hit the top of
the chamber, or downwards to the sonotrode.

\begin{figure}
\begin{center}
\resizebox{7cm}{!}{\includegraphics{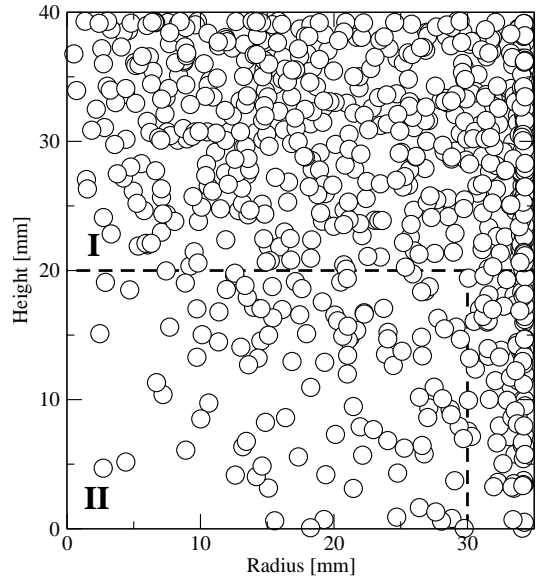}}
\end{center}
\caption{\label{snap} A snapshot of the $N=1000$ shot inside 
the chamber showing the density difference  close to the top and close
to the bottom  wall. An average  has been  performed over the  azimutal
angle, which  explains observed overlaps discs.  For clarity, the size
of the spheres has been reduced. The broken lines serve to define the regions
used in the discussion (see text) and in Fig. \ref{anglecompar}.}
\end{figure}

\begin{figure}
\begin{center}
\resizebox{7cm}{!}{\includegraphics{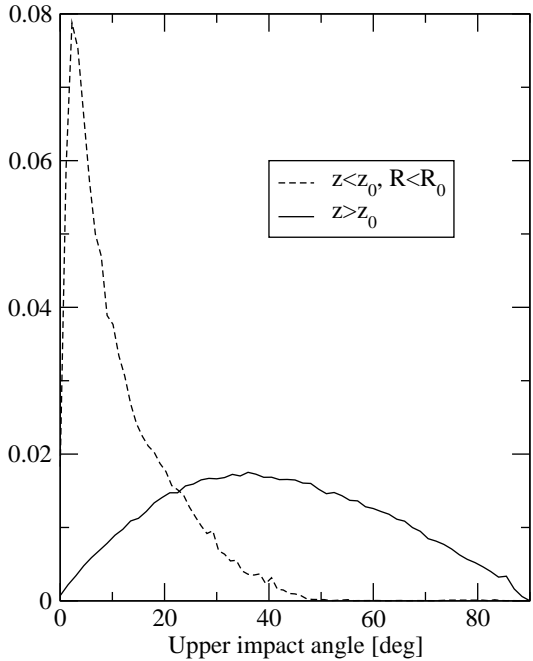}}
\end{center}
\caption{\label{anglecompar} Upper impact angle distribution for spheres
belonging to region I ($z>z_0$, solid line, same as Fig. 
\ref{angletop}) and region II ($z<z_0$, $R<R_0$
broken line) for $N=1000$. Both distributions are normalized to one.}
\end{figure}

This tends to separate the chamber into two parts (Fig. \ref{snap}). 
A first part (at $z>z_0$ with
$z_0\simeq 20~mm$) that connects to the top wall where the density is large
($\eta=8.12×10^{-3}~mm^{-3}$ and a packing fraction of $0.115$). On the other hand, 
the lower part of the chamber corresponds to a much
more dilute situation ($\eta=4.87×10^{-3}~mm^{-3}$ and a packing fraction 
of $0.069$). Consequently, the upper impact angle distribution
is radically different. The impact angle of the spheres bouncing on the more dense
media at $z>z_0$ is very close to $\theta=0^o$. This contrasts with the impact angle 
distribution of the upper part (Fig. \ref{anglecompar}).
\section{Summary and conclusion}
Simulations    on inelastic hard      sphere models using constant
restitution coefficients clearly  show that  the inelasticity of   the
side-wall collisions plays a  key role in  the  impact profile of  the
bottom  and   the top  walls of   a  peening chamber.  With  increased
dissipation, an increased  heterogeneity of the impacts
arising from the accumulation of the  spheres  on the side walls is found.  The
corresponding surface  frequency of  impacts  shows  that  the  latter
effect  is  one  order  of  magnitude  larger  than  the heterogeneity
produced simple by oblique collisions arising from the side walls.

A model  using variable restitution  coefficients permits us to  study in
more detail the effect of the shot  density (or number $N$ of spheres)
in the  chamber. It shows  that  different peening regimes  on the top
wall take place  with changing $N$ that  range from normal impacts for
dilute  granular gases,  to  oblique impacts with  a well-defined mean
impact   velocity.   Densfication close to the  side  walls produces the
occurrence of a local order at a  distance of about the
sphere diameter from the side walls.

These results  suggest that elastic  control  of the  side wall  and a
careful selection  of the  shot density  will permit us  to tune  peening
regimes for the shot and allow various kinds of surface treatments.
The high density observed in the vicinity of the side walls is associated 
with a downwards helicoidal trajectory of the spheres. We believe that
the use of a very rough surface on the side wall could lead to the
reinjection of the spheres in the bulk. Further consideration in this
direction along with deeper experimental characterization of the
heterogeneity is currently under consideration.

Acknowledgments:  Ongoing discussions and collaboration with 
Julian Talbot is gratefully acknowledged. LPTMC is Unité Mixte de Recherche
du CNRS n. 7600. LASMIS is FRE CNRS 2719.
%\bibliography{cooling}

\end{document}